\documentclass[9pt,a4paper,english]{IEEEtran}
\usepackage[T1]{fontenc}
\usepackage{verbatim}
\usepackage{float}

\usepackage{amstext}
\usepackage{graphicx}

\makeatletter

\floatstyle{ruled}
\newfloat{algorithm}{tbp}{loa}
\providecommand{\algorithmname}{Algorithm}
\floatname{algorithm}{\protect\algorithmname}

\usepackage{amsmath}
\usepackage{amssymb}
\usepackage[numbers,sort]{natbib}

\usepackage{graphicx}
\usepackage{subfigure}
\usepackage{pdfsync}
\usepackage{url}
\usepackage{pdfsync}
\usepackage{todonotes}

\makeatother

\usepackage{babel}
\begin{document}
\pagenumbering{gobble}

\title{A Parallel Best-Response Algorithm with Exact Line Search for Nonconvex Sparsity-Regularized Rank Minimization}
\author{\IEEEauthorblockN{Yang Yang$^\dagger$ and Marius Pesavento$^\ddagger$\\ \footnote{helo}} \IEEEauthorblockA{$\dagger$ University of Luxembourg, yang.yang@uni.lu }\\\IEEEauthorblockA{$\ddagger$ Technische Universit\"{a}t Darmstadt, pesavento@nt.tu-darmstadt.de}}\maketitle
\begin{abstract}
In this paper, we propose a convergent parallel best-response algorithm
with the exact line search for the nondifferentiable nonconvex sparsity-regularized
rank minimization problem. On the one hand, it exhibits a faster convergence
than subgradient algorithms and block coordinate descent algorithms.
On the other hand, its convergence to a stationary point is guaranteed,
while ADMM algorithms only converge for convex problems. Furthermore,
the exact line search procedure in the proposed algorithm is performed
efficiently in closed-form to avoid the meticulous choice of stepsizes,
which is however a common bottleneck in subgradient algorithms and
successive convex approximation algorithms. Finally, the proposed
algorithm is numerically tested.\end{abstract}


\section{Introduction}

In this paper, we consider the estimation of a low rank matrix $\mathbf{X}\in\mathbb{R}^{N\times K}$
and a sparse matrix $\mathbf{S}\in\mathbb{R}^{I\times K}$ from noisy
measurements $\mathbf{Y}\in\mathbb{R}^{N\times K}$ such that
\[
\mathbf{Y}=\mathbf{X}+\mathbf{DS}+\mathbf{V},
\]
where $\mathbf{D}\in\mathbb{R}^{N\times I}$ is a known matrix. The
rank of $\mathbf{X}$ is much smaller than $N$ and $K$, i.e, $\textrm{rank}(\mathbf{X})\ll\min(N,K)$,
and the support size of $\mathbf{S}$ is much smaller than $IK$,
i.e., $\left\Vert \mathbf{S}\right\Vert _{0}\ll IK$.

A natural measure for the data mismatch is the least square error
augmented by regularization functions to promote the rank sparsity
of $\mathbf{X}$ and support sparsity of $\mathbf{S}$:
\begin{align*}
(\textrm{SRRM}):\underset{\mathbf{X},\mathbf{S}}{\textrm{minimize}}\quad & \frac{1}{2}\left\Vert \mathbf{X}+\mathbf{D}\mathbf{S}-\mathbf{Y}\right\Vert _{F}^{2}+\frac{\lambda}{2}\left\Vert \mathbf{X}\right\Vert _{*}+\mu\left\Vert \mathbf{S}\right\Vert _{1},
\end{align*}
where $\left\Vert \mathbf{X}\right\Vert _{*}$ is the nuclear norm
of $\mathbf{X}$. This sparsity-regularized rank minimization (SRRM)
problem plays a fundamental role in the analysis of traffic anomalies
in large-scale backbone networks \cite{Mardani2013a}. In this application,
$\mathbf{X}=\mathbf{R}\mathbf{Z}$ where $\mathbf{Z}$ is the unknown
traffic flows over the time horizon of interest, $\mathbf{R}$ is
a given fat routing matrix, $\mathbf{S}$ is the traffic volume anomalies.
The matrix $\mathbf{X}$ inherits the rank sparsity from $\mathbf{Z}$
because common temporal patterns among the traffic flows in addition
to their periodic behavior render most rows/columns of $\mathbf{Z}$
linearly dependent and thus low rank, and $\mathbf{S}$ is assumed
to be sparse because traffic anomalies are expected to happen sporadically
and last shortly relative to the measurement interval, which is represented
by the number of columns $K$.

Although problem (SRRM) is convex, it cannot be easily solved by standard
solvers when the problem dimension is large, for the reason that the
nuclear norm $\left\Vert \mathbf{X}\right\Vert _{*}$ is neither differentiable
nor decomposable among the blocks of $\mathbf{X}$. It follows from
the fact \cite{Recht2010,Steffens2017}
\[
\left\Vert \mathbf{X}\right\Vert _{*}=\min_{(\mathbf{P},\mathbf{Q})}\frac{1}{2}\left(\left\Vert \mathbf{P}\right\Vert _{F}^{2}+\left\Vert \mathbf{Q}\right\Vert _{F}^{2}\right),\textrm{ s.t. }\mathbf{P}\mathbf{Q}=\mathbf{X}
\]
that it may be useful to consider the following optimization problem
where the nuclear norm $\left\Vert \mathbf{X}\right\Vert _{*}$ is
replaced by $\left\Vert \mathbf{P}\right\Vert _{F}^{2}+\left\Vert \mathbf{Q}\right\Vert _{F}^{2}$:
\begin{equation}
\underset{\mathbf{P},\mathbf{Q},\mathbf{S}}{\textrm{minimize}}\;\frac{1}{2}\left\Vert \mathbf{P}\mathbf{Q}+\mathbf{D}\mathbf{S}-\mathbf{Y}\right\Vert _{F}^{2}+\frac{\lambda}{2}\left(\left\Vert \mathbf{P}\right\Vert _{F}^{2}+\left\Vert \mathbf{Q}\right\Vert _{F}^{2}\right)+\mu\left\Vert \mathbf{S}\right\Vert _{1},\label{eq:problem-formulation}
\end{equation}
where $\mathbf{P}\in\mathbb{R}^{N\times\rho}$ and $\mathbf{Q}\in\mathbb{R}^{\rho\times K}$
for a $\rho$ that is usually much smaller than $N$ and $K$: $\rho\ll\min(N,K)$.
Despite the fact that problem (\ref{eq:problem-formulation}) is nonconvex,
it is shown in \cite[Prop. 1]{Mardani2013} that every stationary
point of (\ref{eq:problem-formulation}) is a global optimal solution
of (SRRM) under some mild conditions.

A block coordinate descent (BCD) algorithm is adapted in \cite{Mardani2013b}
to find a stationary point of the nonconvex problem (\ref{eq:problem-formulation}).
In the BCD algorithm, the variables are updated in a cyclic order.
For example, when $\mathbf{P}$ (or $\mathbf{Q}$) is updated, the
variables $(\mathbf{Q,S})$ (or $(\mathbf{P,S})$) are fixed. However,
when fixing $(\mathbf{P,Q})$ and updating $\mathbf{S}$, the elements
of $\mathbf{S}$ are updated element-wise in a sequential order to
reduce the complexity. This is because the optimization problem w.r.t.
$s_{i,k}$, the $(i,k)$-th element of $\mathbf{S}$, has a closed-form
solution:
\[
\underset{s_{i,k}}{\textrm{minimize}}\;\frac{1}{2}\left\Vert \mathbf{P}\mathbf{Q}+\mathbf{D}\mathbf{S}-\mathbf{Y}\right\Vert _{F}^{2}+\frac{\lambda}{2}\left(\left\Vert \mathbf{P}\right\Vert _{F}^{2}+\left\Vert \mathbf{Q}\right\Vert _{F}^{2}\right)+\mu\left\Vert \mathbf{S}\right\Vert _{1},
\]
while the joint optimization problem with respect to (w.r.t.) all
elements of the matrix variable $\mathbf{S}$ does not have a closed-form
solution and is thus not easy to solve. Nevertheless, a drawback of
the sequential element-wise update is that it may incur a large delay
because $s_{i+1,k}$ cannot be updated until $s_{i,k}$ is updated
and the delay may be very large when $I$ is large, which is a norm
rather than an exception in big data analytics \cite{Slavakis2014a}.

The alternating direction method of multipliers (ADMM) algorithm enables
the simultaneous update of all elements of $\mathbf{S}$, but it does
not have a guarantee convergence to a stationary point because the
optimization problem (\ref{eq:problem-formulation}) is nonconvex
\cite{Mardani2013}. Note that there is some recent development in
ADMM for nonconvex problems, see \cite{Hong2016,Jiang2016} for example
and the references therein. The ADMM algorithm proposed in \cite{Hong2016}
is designed for nonconvex sharing/consensus problems, and cannot be
applied to solve problem (\ref{eq:problem-formulation}). The ADMM
algorithm proposed in \cite{Jiang2016} converges if the matrix $\mathbf{D}$
in (\ref{eq:problem-formulation}) has full row rank, which is however
not necessarily the case.

The nondifferentiable nonconvex problem (\ref{eq:problem-formulation})
can also be solved by standard subgradient and/or successive convex
approximation (SCA) algorithms \cite{Scutari_BigData}. However, convergence
of subgradient and SCA algorithms is mostly established under diminishing
stepsizes, which is sometimes difficult to deploy in practice because
the convergence behavior is sensitive to the decay rate. As a matter
of fact, the meticulous choice of stepsizes severely limits the applicability
of subgradient and SCA algorithms in nonsmooth optimization and big
data analytics \cite{Slavakis2014a}.

In this paper, we propose a convergent parallel best-response algorithm,
where all elements of $\mathbf{P}$, $\mathbf{Q}$ and $\mathbf{S}$
are updated simultaneously. This is a well known concept in optimization
and sometimes listed under different names, for example, the parallel
block coordinate descent algorithm (cf. \cite{Elad2006}) and the
Jacobi algorithm (cf. \cite{Scutarib}). To accelerate the convergence,
we compute the stepsize by the exact line search procedure proposed
in \cite{Yang_ConvexApprox}: the exact line search is performed over
a properly designed differentiable function and the resulting stepsize
can be expressed in a closed-form expression, so that the computational
complexity is much lower than the traditional line search which is
over the original nondifferentiable objective function. The proposed
algorithm has several attractive features: i) the variables are updated
simultaneously based on the best-response; ii) the stepsize is computed
in closed-form based on the exact line search; iv) it converges to
a stationary point, and its advantages over existing algorithms are
summarized as follows:
\begin{itemize}
\item Feature i) is an advantage over the BCD algorithm;
\item Features i) and ii) are advantages over subgradient algorithms;
\item Feature ii) is an advantage over SCA algorithms;
\item Feature iii) is an advantage over ADMM algorithms.
\end{itemize}
The above advantages will further be illustrated by numerical results.

\section{The Proposed Parallel Best-Response Algorithm with Exact Line Search}

In this section, we propose an iterative algorithm to find a stationary
point of problem (\ref{eq:problem-formulation}). It consists of solving
a sequence of successively refined approximate problems, which are
presumably much easier to solve than the original problem. To this
end, we define
\begin{align*}
f(\mathbf{P},\mathbf{Q},\mathbf{S}) & \triangleq\frac{1}{2}\left\Vert \mathbf{P}\mathbf{Q}+\mathbf{D}\mathbf{S}-\mathbf{Y}\right\Vert _{F}^{2}+\frac{\lambda}{2}\left(\left\Vert \mathbf{P}\right\Vert _{F}^{2}+\left\Vert \mathbf{Q}\right\Vert _{F}^{2}\right),\\
g(\mathbf{S}) & \triangleq\mu\left\Vert \mathbf{S}\right\Vert _{1}.
\end{align*}

Although $f(\mathbf{P,Q,S})$ in (\ref{eq:problem-formulation}) is
not jointly convex w.r.t. $(\mathbf{P,Q,S})$, it is individual convex
in $\mathbf{P}$, $\mathbf{Q}$ and $\mathbf{S}$. In other words,
$f(\mathbf{P,Q,S})$ is convex w.r.t. one variable while the other
two variables are fixed. Preserving and exploiting this partial convexity
considerably accelerates the convergence and it has become the central
idea in the successive convex approximation and the successive pseudoconvex
approximation \cite{Scutarib,Yang_ConvexApprox}.

To simplify the notation, we use $\mathbf{Z}$ as a compact notation
for $(\mathbf{P},\mathbf{Q},\mathbf{S})$: $\mathbf{Z}\triangleq(\mathbf{P},\mathbf{Q},\mathbf{S})$;
in the rest of the paper, $\mathbf{Z}$ and $(\mathbf{P},\mathbf{Q},\mathbf{S})$
are used interchangeably. Given $\mathbf{Z}^{t}=(\mathbf{P}^{t},\mathbf{Q}^{t},\mathbf{S}^{t})$
in iteration $t$, we approximate the original nonconvex function
$f(\mathbf{Z})$ by a convex function $\tilde{f}(\mathbf{Z};\mathbf{Z}^{t})$
that is of the following form:
\begin{equation}
\tilde{f}(\mathbf{Z};\mathbf{Z}^{t})=\tilde{f}_{P}(\mathbf{P};\mathbf{Z}^{t})+\tilde{f}_{Q}(\mathbf{Q};\mathbf{Z}^{t})+\tilde{f}_{S}(\mathbf{S};\mathbf{Z}^{t}),\label{eq:approximate-function}
\end{equation}
where\begin{subequations}\label{eq:approximate-function-individual}
\begin{align}
\tilde{f}_{P}(\mathbf{P};\mathbf{Z}^{t}) & \triangleq f(\mathbf{P},\mathbf{Q}^{t},\mathbf{S}^{t})\nonumber \\
 & =\frac{1}{2}\left\Vert \mathbf{P}\mathbf{Q}^{t}+\mathbf{D}\mathbf{S}^{t}-\mathbf{Y}\right\Vert _{F}^{2}+\frac{\lambda}{2}\left\Vert \mathbf{P}\right\Vert _{F}^{2},\label{eq:approximate-function-P}\\
\tilde{f}_{Q}(\mathbf{Q};\mathbf{Z}^{t}) & \triangleq f(\mathbf{P}^{t},\mathbf{Q},\mathbf{S}^{t})\nonumber \\
 & =\frac{1}{2}\left\Vert \mathbf{P}^{t}\mathbf{Q}+\mathbf{D}\mathbf{S}^{t}-\mathbf{Y}\right\Vert _{F}^{2}+\frac{\lambda}{2}\left\Vert \mathbf{Q}\right\Vert _{F}^{2},\label{eq:approximate-function-Q}\\
\tilde{f}_{S}(\mathbf{S};\mathbf{Z}^{t}) & \triangleq\sum_{i,k}f(\mathbf{P}^{t},\mathbf{Q}^{t},s_{i,k},(s_{j,k}^{t})_{j\neq i},(\mathbf{s}_{j}^{t})_{j\neq i})\nonumber \\
 & =\sum_{i,k}\frac{1}{2}\left\Vert \mathbf{P}^{t}\mathbf{q}_{k}^{t}+\mathbf{d}_{i}s_{i,k}+\sum_{j\neq i}\mathbf{d}_{j}s_{j,k}^{t}-\mathbf{y}_{k}\right\Vert _{2}^{2}\nonumber \\
 & =\textrm{tr}(\mathbf{S}^{T}\mathbf{d}(\mathbf{D}^{T}\mathbf{D})\mathbf{S})\nonumber \\
 & \qquad-\textrm{tr}(\mathbf{S}^{T}(\mathbf{d}(\mathbf{D}^{T}\mathbf{D})\mathbf{S}^{t}-\mathbf{D}^{T}(\mathbf{D}\mathbf{S}^{t}-\mathbf{Y}+\mathbf{P}^{t}\mathbf{Q}^{t}))),\label{eq:approximate-function-S}
\end{align}
\end{subequations}with $\mathbf{q}_{k}$ (or $\mathbf{y}_{k}$) and
$\mathbf{d}_{i}$ denoting the $k$-th and $i$-th column of $\mathbf{Q}$
(or $\mathbf{Y}$) and $\mathbf{D}$, respectively, while $\mathbf{d}(\mathbf{D}^{T}\mathbf{D})$
denotes a diagonal matrix with elements on the main diagonal identical
to those of the matrix $\mathbf{D}^{T}\mathbf{D}$. Note that in the
approximate function w.r.t. $\mathbf{P}$ and $\mathbf{Q}$, the remaining
variables $(\mathbf{Q,S})$ and $(\mathbf{P,S})$ are fixed, respectively.
Although it is tempting to define the approximate function of $f(\mathbf{P,Q,S})$
w.r.t. $\mathbf{S}$ by fixing $\mathbf{P}$ and $\mathbf{Q}$, minimizing
$f(\mathbf{P}^{t},\mathbf{Q}^{t},\mathbf{S})$ w.r.t. the matrix variable
$\mathbf{S}$ does not have a closed-form solution and must be solved
iteratively. Therefore the proposed approximate function $\tilde{f}_{S}(\mathbf{S};\mathbf{Z}^{t})$
in (\ref{eq:approximate-function-S}) consists of $IK$ component
functions, and in the $(i,k)$-th component function, $s_{i,k}$ is
the variable while all other variables are fixed, namely, $\mathbf{P}$,
$\mathbf{Q}$, $(s_{j,k})_{j\neq i}$, and $(\mathbf{s}_{j})_{j\neq i}$.
As we will show shortly, minimizing $\tilde{f}(\mathbf{S};\mathbf{Z}^{t})$
w.r.t. $\mathbf{S}$ exhibits a closed-form solution.

We remark that the approximate function $\tilde{f}(\mathbf{Z;Z}^{t})$
is a (strongly) convex function and it is differentiable in both $\mathbf{Z}$
and $\mathbf{Z}^{t}$. Furthermore, the gradient of the approximate
function $\tilde{f}(\mathbf{P,Q,S};\mathbf{Z}^{t})$ is equal to that
of $f(\mathbf{P,Q,S})$ at $\mathbf{Z}=\mathbf{Z}^{t}$. To see this:

\begin{align*}
\nabla_{\mathbf{P}}\tilde{f}(\mathbf{Z};\mathbf{Z}^{t}) & =\nabla_{\mathbf{P}}\tilde{f}_{P}(\mathbf{P};\mathbf{Z}^{t})\\
 & =\nabla_{\mathbf{P}}\left.\left(\frac{1}{2}\left\Vert \mathbf{P}\mathbf{Q}^{t}+\mathbf{D}\mathbf{S}^{t}-\mathbf{Y}\right\Vert _{F}^{2}+\frac{\lambda}{2}\left\Vert \mathbf{P}\right\Vert _{F}^{2}\right)\right|_{\mathbf{P=P}^{t}}\\
 & =\nabla_{\mathbf{P}}\left.f(\mathbf{P,Q,S})\right|_{\mathbf{Z=Z}^{t}},
\end{align*}
\begin{align*}
\nabla_{\mathbf{Q}}\tilde{f}(\mathbf{Z};\mathbf{Z}^{t}) & =\nabla_{\mathbf{Q}}\tilde{f}_{Q}(\mathbf{Q};\mathbf{Z}^{t})\\
 & =\nabla_{\mathbf{Q}}\left.\left(\frac{1}{2}\left\Vert \mathbf{P}^{t}\mathbf{Q}+\mathbf{D}\mathbf{S}^{t}-\mathbf{Y}\right\Vert _{F}^{2}+\frac{\lambda}{2}\left\Vert \mathbf{Q}\right\Vert _{F}^{2}\right)\right|_{\mathbf{Q=Q}^{t}}\\
 & =\left.\nabla_{\mathbf{Q}}f(\mathbf{P,Q,S})\right|_{\mathbf{Z}=\mathbf{Z}^{t}}.
\end{align*}
and $\nabla_{\mathbf{S}}\tilde{f}(\mathbf{Z};\mathbf{Z}^{t})=(\nabla_{s_{i,k}}\tilde{f}(\mathbf{Z};\mathbf{Z}^{t}))_{i,k}$
while
\begin{align*}
\nabla_{s_{i,k}}\tilde{f}(\mathbf{Z};\mathbf{Z}^{t}) & =\nabla_{s_{i,k}}\tilde{f}_{S}(\mathbf{S};\mathbf{Z}^{t})\\
 & =\nabla_{s_{i,k}}f(\mathbf{P}^{t},\mathbf{Q}^{t},s_{i,k},\mathbf{s}_{i,-k}^{t},\mathbf{s}_{-i}^{t})\\
 & =\left.\nabla_{s_{i,k}}f(\mathbf{P,Q,S})\right|_{\mathbf{Z}=\mathbf{Z}^{t}}.
\end{align*}

In iteration $t$, the approximate problem consists of minimizing
the approximate function over the same feasible set as the original
problem (\ref{eq:problem-formulation}):
\begin{align}
\underset{\mathbf{Z}=(\mathbf{P},\mathbf{Q},\mathbf{S})}{\textrm{minimize}} & \quad\tilde{f}(\mathbf{Z};\mathbf{Z}^{t})+g(\mathbf{S}).\label{eq:approximate-problem}
\end{align}
Since $\tilde{f}(\mathbf{Z};\mathbf{Z}^{t})$ is strongly convex in
$\mathbf{Z}$ and $g(\mathbf{S})$ is a convex function w.r.t. $\mathbf{S}$,
the approximate problem (\ref{eq:approximate-problem}) is convex
and it has a unique (globally) optimal solution, which is denoted
as $\mathbb{B}\mathbf{Z}^{t}=(\mathbb{B}_{P}\mathbf{Z}^{t},\mathbb{B}_{Q}\mathbf{Z}^{t},\mathbb{B}_{S}\mathbf{Z}^{t})$.

The approximate problem (\ref{eq:approximate-problem}) naturally
decomposes into several smaller problems which can be solved in parallel:\begin{subequations}\label{eq:approximate-problem-solution}
\begin{align}
\mathbb{B}_{P}\mathbf{Z}^{t} & \triangleq\underset{\mathbf{P}_{k}}{\arg\min}\;\tilde{f}_{P}(\mathbf{P};\mathbf{Z}^{t})\nonumber \\
 & =(\mathbf{Y}-\mathbf{D}\mathbf{S}^{t})(\mathbf{Q}^{t})^{T}(\mathbf{Q}^{t}(\mathbf{Q}^{t})^{T}+\lambda\mathbf{I})^{-1},\label{eq:BP}
\end{align}
\begin{align}
\mathbb{B}_{Q}\mathbf{Z}^{t} & \triangleq\underset{\mathbf{Q}}{\arg\min}\;\tilde{f}_{Q}(\mathbf{Q};\mathbf{Z}^{t})\nonumber \\
 & =((\mathbf{P}^{t})^{T}\mathbf{P}^{t}+\lambda\mathbf{I})^{-1}(\mathbf{P}^{t})^{T}(\mathbf{Y}-\mathbf{D}\mathbf{S}^{t}),\label{eq:BQ}
\end{align}
\begin{align}
\mathbb{B}_{S}\mathbf{Z}^{t} & \triangleq\underset{\mathbf{S}}{\arg\min}\;\tilde{f}_{S}(\mathbf{S};\mathbf{Z}^{t})+g(\mathbf{S})\nonumber \\
 & =\mathbf{d}(\mathbf{D}^{T}\mathbf{D})^{-1}\cdot\nonumber \\
 & \qquad\mathcal{S}_{\mu}\left(\mathbf{d}(\mathbf{D}^{T}\mathbf{D})\mathbf{S}^{t}-\mathbf{D}^{T}(\mathbf{D}\mathbf{S}^{t}-\mathbf{Y}^{t}+\mathbf{P}^{t}\mathbf{Q}^{t})\right),\label{eq:BS}
\end{align}
\end{subequations}where $\mathcal{S}_{\mu}(\mathbf{X})$ is an element-wise
soft-threshold operator: the $(i,j)$-th element of $\mathcal{S}_{\mu}(\mathbf{X})$
is $[X_{ij}-\lambda]^{+}-[-X_{ij}-\lambda]^{+}$. As we can readily
see from (\ref{eq:approximate-problem-solution}), the approximate
problems can be solved efficiently because the optimal solutions are
provided in an analytical expression.

Since $\tilde{f}(\mathbf{Z};\mathbf{Z}^{t})$ is convex in $\mathbf{Z}$
and differentiable in both $\mathbf{Z}$ and $\mathbf{Z}^{t}$, and
has the same gradient as $f(\mathbf{Z})$ at $\mathbf{Z}=\mathbf{Z}^{t}$,
it follows from \cite[Prop. 1]{Yang_ConvexApprox} that $\mathbb{B}\mathbf{Z}^{t}-\mathbf{Z}^{t}$
is a descent direction of the original objective function $f(\mathbf{Z})+g(\mathbf{S})$
at $\mathbf{Z}=\mathbf{Z}^{t}$. The variable update in the $t$-th
iteration is thus defined as follows:\begin{subequations}\label{eq:variable-update}
\begin{align}
\mathbf{P}^{t+1} & =\mathbf{P}^{t}+\gamma(\mathbb{B}_{P}\mathbf{Z}^{t}-\mathbf{P}^{t}),\label{eq:variable-update-P}\\
\mathbf{Q}^{t+1} & =\mathbf{Q}^{t}+\gamma(\mathbb{B}_{Q}\mathbf{Z}^{t}-\mathbf{Q}^{t}),\label{eq:variable-update-Q}\\
\mathbf{S}^{t+1} & =\mathbf{S}^{t}+\gamma(\mathbb{B}_{S}\mathbf{Z}^{t}-\mathbf{S}^{t}),\label{eq:variable-update-S}
\end{align}
\end{subequations}where $\gamma\in(0,1]$ is the stepsize that should
be properly selected.

A natural (and traditional) choice of the stepsize $\gamma$ is given
by the exact line search:
\begin{equation}
\min_{0\leq\gamma\leq1}\left\{ f(\mathbf{Z}^{t}+\gamma(\mathbb{B}\mathbf{Z}^{t}-\mathbf{Z}^{t}))+g(\mathbf{S}^{t}+\gamma(\mathbb{B}_{S}\mathbf{Z}^{t}-\mathbf{S}^{t}))\right\} ,\label{eq:line-search-traditional}
\end{equation}
in which the stepsize that yields the largest decrease in objective
function value along the direction $\mathbb{B}\mathbf{Z}^{t}-\mathbf{Z}^{t}$
is selected. Nevertheless, this choice leads to high computational
complexity, because $g(\mathbf{S})$ is nondifferentiable and the
exact line search involves minimizing a nondifferentiable function.
Alternatives include constant stepsizes and diminishing stepsizes.
However, they suffer from slow convergence (cf. \cite{Scutarib})
and parameter tuning (cf. \cite{Yang_ConvexApprox}). As a matter
of fact, the meticulous choice of stepsizes have become a major bottleneck
for subgradient and successive convex approximation algorithm \cite{Slavakis2014a}.

It is shown in \cite[Sec. III-A]{Yang_ConvexApprox} that to achieve
convergence, it suffices to perform the exact line search over the
following differentiable function:
\begin{equation}
f(\mathbf{Z}^{t}+\gamma(\mathbb{B}\mathbf{Z}^{t}-\mathbf{Z}^{t}))+g(\mathbf{S}^{t})+\gamma(g(\mathbb{B}_{S}\mathbf{Z}^{t})-g(\mathbf{S}^{t})),\label{eq:line-search-proposed-concept}
\end{equation}
which is an upper bound of the objective function in (\ref{eq:line-search-traditional})
after applying Jensen's inequality to the convex nondifferentiable
function $g(\mathbf{S})$:
\begin{align*}
g(\mathbf{S}^{t}+\gamma(\mathbb{B}_{S}\mathbf{Z}^{t}-\mathbf{S}^{t})) & \leq g(\mathbf{S}^{t})+\gamma(g(\mathbb{B}_{S}\mathbf{Z}^{t})-g(\mathbf{S}^{t})).
\end{align*}
This exact line search procedure over the differentiable function
(\ref{eq:line-search-proposed-concept}) achieves a good tradeoff
between performance and complexity. Furthermore, after substituting
the expressions of $f(\mathbf{Z})$ and $g(\mathbf{S})$ into (\ref{eq:line-search-proposed-concept}),
the exact line search boils down to minimizing a four order polynomial
over the interval $[0,1]$:
\begin{align}
\gamma^{t} & =\underset{0\leq\gamma\leq1}{\arg\min}\left\{ f(\mathbf{Z}^{t}+\gamma(\mathbb{B}\mathbf{Z}^{t}-\mathbf{Z}^{t}))+\gamma(g(\mathbb{B}_{S}\mathbf{X}^{t})-g(\mathbf{S}^{t}))\right\} \nonumber \\
 & =\underset{0\leq\gamma\leq1}{\arg\min}\left\{ \frac{1}{4}a\gamma^{4}+\frac{1}{3}b\gamma^{3}+\frac{1}{2}c\gamma^{2}+d\gamma\right\} ,\label{eq:line-search-proposed}
\end{align}
where
\begin{align*}
a & \triangleq2\left\Vert \triangle\mathbf{P}^{t}\triangle\mathbf{Q}^{t}\right\Vert _{F}^{2},\\
b & \triangleq3\textrm{tr}(\triangle\mathbf{P}^{t}\triangle\mathbf{Q}^{t}(\mathbf{P}^{t}\triangle\mathbf{Q}^{t}+\triangle\mathbf{P}^{t}\mathbf{Q}^{t}+\mathbf{D}\triangle\mathbf{S}^{t})^{T}),\\
c & \triangleq2\textrm{tr}(\triangle\mathbf{P}^{t}\triangle\mathbf{Q}^{t}(\mathbf{P}^{t}\mathbf{Q}^{t}+\mathbf{D}\mathbf{S}^{t}-\mathbf{Y}^{t})^{T})\\
 & \quad+\left\Vert \mathbf{P}^{t}\triangle\mathbf{Q}^{t}+\triangle\mathbf{P}^{t}\mathbf{Q}^{t}+\mathbf{D}\triangle\mathbf{S}^{t}\right\Vert _{F}^{2}\\
 & \quad+\lambda(\left\Vert \triangle\mathbf{P}^{t}\right\Vert _{F}^{2}+\left\Vert \triangle\mathbf{Q}^{t}\right\Vert _{F}^{2}),\\
d & \triangleq\textrm{tr}((\mathbf{P}^{t}\triangle\mathbf{Q}^{t}+\triangle\mathbf{P}^{t}\mathbf{Q}^{t}+\mathbf{D}\triangle\mathbf{S}^{t})(\mathbf{P}^{t}\mathbf{Q}^{t}+\mathbf{D}\mathbf{S}^{t}-\mathbf{Y}^{t}))\\
 & \quad+\lambda(\textrm{tr}(\mathbf{P}^{t}\triangle\mathbf{P}^{t})+\textrm{tr}(\mathbf{Q}^{t}\triangle\mathbf{Q}^{t}))+\mu(\left\Vert \mathbb{B}_{S}\mathbf{X}^{t}\right\Vert _{1}-\left\Vert \mathbf{S}^{t}\right\Vert _{1}),
\end{align*}
for $\triangle\mathbf{P}^{t}\triangleq\mathbb{B}_{P}\mathbf{Z}^{t}-\mathbf{P}^{t}$,
$\triangle\mathbf{Q}^{t}\triangleq\mathbb{B}_{Q}\mathbf{Z}^{t}-\mathbf{Q}^{t}$
and $\triangle\mathbf{S}^{t}\triangleq\mathbb{B}_{S}\mathbf{Z}^{t}-\mathbf{S}^{t}$.
Finding the optimal points of (\ref{eq:line-search-proposed}) is
equivalent to finding the nonnegative real root of a third-order polynomial.
Making use of Cardano's method, we could express $\gamma^{t}$ defined
in (\ref{eq:line-search-proposed}) in a closed-form expression:\begin{subequations}\label{eq:line-search-proposed-closed-form}
\begin{align}
\gamma^{t} & =[\bar{\gamma}^{t}]_{0}^{1},\label{eq:line-search-proposed-closed-form-1}\\
\bar{\gamma}^{t} & =\sqrt[3]{\Sigma_{1}+\sqrt{\Sigma_{1}^{2}+\Sigma_{2}^{3}}}+\sqrt[3]{\Sigma_{1}-\sqrt{\Sigma_{1}^{2}+\Sigma_{2}^{3}}}-\frac{b}{3a},\label{eq:line-search-proposed-closed-form-2}
\end{align}
\end{subequations}where $\left[\bar{\gamma}^{t}\right]_{0}^{1}=\max(\min(\bar{\gamma}^{t},1),0)$
is the projection of $\bar{\gamma}^{t}$ onto the interval $[0,1]$,
$\Sigma_{1}\triangleq-(b/3a)^{3}+bc/6a^{2}-d/2a$ and $\Sigma_{2}\triangleq c/3a-(b/3a)^{2}$.
Note that in (\ref{eq:line-search-proposed-closed-form-2}), the right
hand side has three values (two of them could be complex numbers),
and the equal sign reads to be equal to the smallest one among the
real nonnegative values.

\begin{algorithm}[t]
\textbf{Data: }$t=0$, $\mathbf{Z}^{0}$ (arbitrary but fixed, e.g.,
$\mathbf{Z}^{0}=\mathbf{0}$), stop criterion $\delta$.

\textbf{S1: }Compute $(\mathbb{B}_{P}\mathbf{Z}^{t},\mathbb{B}_{Q}\mathbf{Z}^{t},\mathbb{B}_{S}\mathbf{Z}^{t})$
according to (\ref{eq:approximate-problem-solution}).

\textbf{S2: }Determine the stepsize $\gamma^{t}$ by the exact line
search (\ref{eq:line-search-proposed-closed-form}).

\textbf{S3: }Update $(\mathbf{P},\mathbf{Q},\mathbf{Z})$ according
to (\ref{eq:variable-update}).

\textbf{S4: }If $\left|\textrm{tr}((\mathbb{B}\mathbf{Z}^{t}-\mathbf{Z}^{t})^{T}\nabla f(\mathbf{Z}^{t}))\right|\leq\delta$,
STOP; otherwise go to \textbf{S1}.

\protect\caption{\label{alg:Successive-approximation-method}The parallel best-response
algorithm with exact line search for problem (\ref{eq:problem-formulation})}
\end{algorithm}

The proposed algorithm is summarized in Algorithm \ref{alg:Successive-approximation-method},
and we draw a few comments on its attractive features and advantages.

\textbf{On the parallel best-response update:} In each iteration,
the variables $\mathbf{P}$, $\mathbf{Q}$, and $\mathbf{S}$ are
updated simultaneously based on the best-response. The improvement
in convergence speed w.r.t. the BCD algorithm in \cite{Mardani2013b}
is notable because in the BCD algorithm, the optimization w.r.t. each
element of $\mathbf{S}$, say $s_{i,k}$, is implemented in a sequential
order, and the number of elements, $IK$, is usually very large in
big data applications. To avoid the meticulous choice of stepsizes
and further accelerate the convergence, the exact line search is performed
over the differentiable function $f(\mathbf{Z}^{t}+\gamma(\mathbb{B}\mathbf{Z}^{t}-\mathbf{Z}^{t}))+\gamma(g(\mathbb{B}_{S}\mathbf{Z}^{t})-g_{S}(\mathbf{Z}^{t}))$
and it can be computed by a closed-form expression. The yields easier
implementation and faster convergence than subgradient and SCA algorithms
with diminishing stepsizes.

\textbf{On the complexity:} The complexity of the proposed algorithm
is maintained at a very low level, because both the best-responses
$(\mathbb{B}_{P}\mathbf{Z}^{t},\mathbb{B}_{Q}\mathbf{Z}^{t},\mathbb{B}_{S}\mathbf{Z}^{t})$
and the exact line search can be computed by closed-form expressions,
cf. (\ref{eq:approximate-function-individual}) and (\ref{eq:line-search-proposed-closed-form}).
Only basic linear algebraic operations are required, reducing the
requirements on the hardware's computational capabilities.

\textbf{On the convergence:} The proposed Algorithm \ref{alg:Successive-approximation-method}
has a guaranteed convergence in the sense that every limit point of
the sequence $\{\mathbf{Z}^{t}\}_{t}$ is a stationary point of problem
(\ref{eq:problem-formulation}). This claim directly follows from
\cite[Theorem 1]{Yang_ConvexApprox}, and it serves as a certificate
for the solution quality compared with ADMM algorithms.

\subsection{Decomposition of the Proposed Algorithm}

The proposed Algorithm \ref{alg:Successive-approximation-method}
can be further decomposed to enable the parallel processing over a
number of $L$ nodes in a distributed network. To see this, we first
decompose the matrix variables $\mathbf{P}$, $\mathbf{D}$ and $\mathbf{Y}$
into multiple blocks $(\mathbf{P}_{l})_{l=1}^{L}$, $(\mathbf{D}_{l})_{l=1}^{L}$
and $(\mathbf{Y}_{l})_{l=1}^{L}$, while $\mathbf{P}_{l}\in\mathbb{R}^{N_{l}\times\rho}$,
$\mathbf{D}_{l}\in\mathbb{R}^{N_{l}\times I}$ and $\mathbf{Y}_{l}\in\mathbb{R}^{N_{l}\times K}$
consists of $N_{l}$ rows of $\mathbf{P}$, $\mathbf{D}$ and $\mathbf{Y}$,
respectively:
\[
\mathbf{P}=\left[\begin{array}{c}
\mathbf{P}_{1}\\
\mathbf{P}_{2}\\
\vdots\\
\mathbf{P}_{L}
\end{array}\right],\mathbf{D}=\left[\begin{array}{c}
\mathbf{D}_{1}\\
\mathbf{D}_{2}\\
\vdots\\
\mathbf{D}_{L}
\end{array}\right],\mathbf{Y}=\left[\begin{array}{c}
\mathbf{Y}_{1}\\
\mathbf{Y}_{2}\\
\vdots\\
\mathbf{Y}_{L}
\end{array}\right],
\]
where each node $l$ has access to the variables $(\mathbf{P}_{l},\mathbf{Q},\mathbf{S})$.
The computation of $\mathbb{B}_{P}\mathbf{Z}^{t}$ in (\ref{eq:variable-update-P})
can be decomposed as $\mathbb{B}_{P}\mathbf{Z}^{t}=(\mathbb{B}_{P,l}\mathbf{Z}^{t})_{l=1}^{L}$:
\[
\mathbb{B}_{P,l}\mathbf{Z}^{t}=(\mathbf{Y}_{l}-\mathbf{D}_{l}\mathbf{S}^{t})(\mathbf{Q}^{t})^{T}(\mathbf{Q}^{t}(\mathbf{Q}^{t})^{T}+\lambda\mathbf{I})^{-1},l=1,\ldots,L.
\]
Accordingly, the computation of $\mathbb{B}_{Q}\mathbf{Z}^{t}$ and
$\mathbb{B}_{S}\mathbf{Z}^{t}$ in (\ref{eq:variable-update-Q}) and
(\ref{eq:variable-update-S}) can be rewritten as
\begin{align*}
\mathbb{B}_{Q}\mathbf{Z}^{t} & =\left({\textstyle \sum_{l=1}^{L}}(\mathbf{P}_{l}^{t})^{T}\mathbf{P}_{l}^{t}+\lambda\mathbf{I}\right)^{-1}\left({\textstyle \sum_{l=1}^{L}}(\mathbf{P}_{l}^{t})^{T}(\mathbf{Y}_{l}-\mathbf{D}_{l}\mathbf{S}^{t})\right),\\
\mathbb{B}_{S}\mathbf{Z}^{t} & =\mathbf{d}\left({\textstyle \sum_{l=1}^{L}}\mathbf{D}_{l}^{T}\mathbf{D}_{l}\right)^{-1}\cdot\\
\mathcal{S}_{\mu} & \left(\mathbf{d}\left({\textstyle \sum_{l=1}^{L}}\mathbf{D}_{l}^{T}\mathbf{D}_{l}\right)\mathbf{S}^{t}-{\textstyle \sum_{l=1}^{L}}\mathbf{D}_{l}^{T}(\mathbf{D}_{l}\mathbf{S}^{t}-\mathbf{Y}_{l}^{t}+\mathbf{P}_{l}^{t}\mathbf{Q}^{t})\right).
\end{align*}
Before determining the stepsize, the computation of $a$ in (\ref{eq:line-search-proposed-closed-form})
can also be decomposed among the nodes as $a=\sum_{l=1}^{L}a_{l}$,
where
\[
a_{l}\triangleq2\left\Vert \triangle\mathbf{P}_{l}^{t}\triangle\mathbf{Q}^{t}\right\Vert _{F}^{2}.
\]
The decomposition of $b$, $c$, and $d$ is similar to that of $a$,
where
\begin{align*}
b_{l} & \triangleq3\textrm{tr}(\triangle\mathbf{P}_{l}^{t}\triangle\mathbf{Q}^{t}(\mathbf{P}_{l}^{t}\triangle\mathbf{Q}^{t}+\triangle\mathbf{P}_{l}^{t}\mathbf{Q}^{t}+\mathbf{D}_{l}\triangle\mathbf{S}^{t})^{T}),\\
c_{l} & \triangleq2\textrm{tr}(\triangle\mathbf{P}_{l}^{t}\triangle\mathbf{Q}^{t}(\mathbf{P}_{l}^{t}\mathbf{Q}^{t}+\mathbf{D}_{l}\mathbf{S}^{t}-\mathbf{Y}_{l}^{t})^{T})\\
 & \quad+\left\Vert \mathbf{P}_{l}^{t}\triangle\mathbf{Q}^{t}+\triangle\mathbf{P}_{l}^{t}\mathbf{Q}^{t}+\mathbf{D}_{l}\triangle\mathbf{S}^{t}\right\Vert _{F}^{2}\\
 & \quad+\lambda\left\Vert \triangle\mathbf{P}_{l}^{t}\right\Vert _{F}^{2}+\frac{\lambda}{I}\left\Vert \triangle\mathbf{Q}_{l}^{t}\right\Vert _{F}^{2},\\
d_{l} & \triangleq\textrm{tr}((\mathbf{P}_{l}^{t}\triangle\mathbf{Q}^{t}+\triangle\mathbf{P}_{l}^{t}\mathbf{Q}^{t}+\mathbf{D}_{l}\triangle\mathbf{S}^{t})(\mathbf{P}_{l}^{t}\mathbf{Q}^{t}+\mathbf{D}_{l}\mathbf{S}^{t}-\mathbf{Y}_{l}^{t}))\\
 & \quad+\lambda\textrm{tr}(\mathbf{P}_{l}^{t}\triangle\mathbf{P}_{l}^{t})+\frac{\lambda}{I}\textrm{tr}(\mathbf{Q}^{t}\triangle\mathbf{Q}^{t})+\frac{\mu}{I}(\left\Vert \mathbb{B}_{S}\mathbf{X}^{t}\right\Vert _{1}-\left\Vert \mathbf{S}^{t}\right\Vert _{1}).
\end{align*}
To compute the stepsize as in (\ref{eq:line-search-proposed-closed-form}),
the nodes mutually exchange $(a_{l},b_{l},c_{l},d_{l})$. The four
dimensional vector $(a_{l},b_{l},c_{l},d_{l})$ provides each node
with all the necessary information to individually calculate $(a,b,c,d)$
and $(\Sigma_{1},\Sigma_{2},\Sigma_{3})$, and then the stepsize $\gamma^{t}$
according to (\ref{eq:line-search-proposed-closed-form}). The signaling
incurred by the exact line search is thus small and affordable.

\section{Numerical Simulations}

\begin{figure}[t]
\center \includegraphics[scale=0.5]{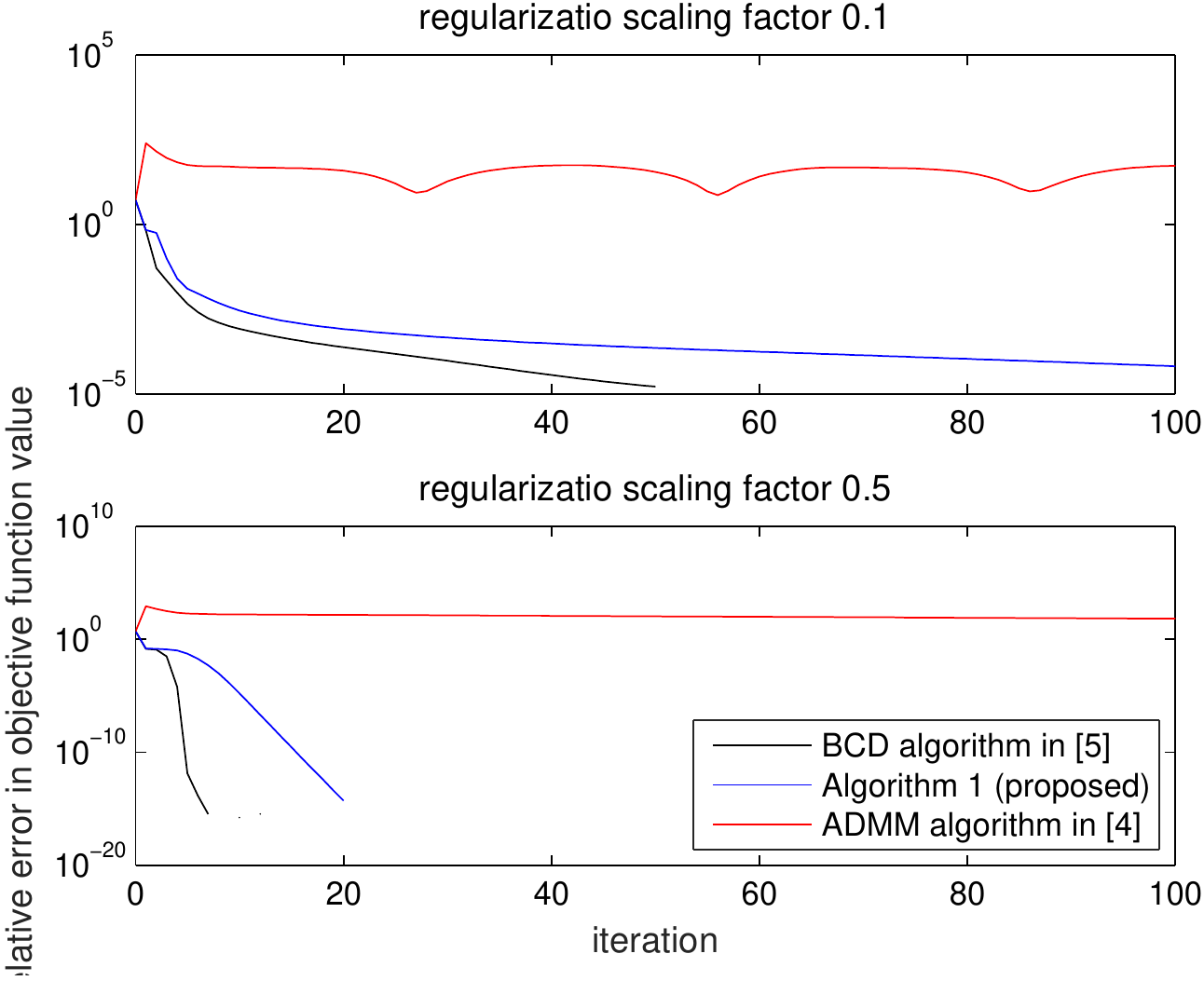}\protect\caption{\label{fig:Iteration}Relative error in objective function value versus
iterations}
\end{figure}

In this section, we perform numerical tests to compare the proposed
Algorithm \ref{alg:Successive-approximation-method} with the BCD
algorithm proposed in \cite{Mardani2013b} and the ADMM algorithm
proposed in \cite{Mardani2013}. We start with a brief description
of the ADMM algorithm: the problem (\ref{eq:problem-formulation})
can be rewritten as
\begin{align}
\underset{\mathbf{P},\mathbf{Q},\mathbf{A},\mathbf{B}}{\textrm{minimize}}\; & \frac{1}{2}\left\Vert \mathbf{P}\mathbf{Q}+\mathbf{D}\mathbf{A}-\mathbf{Y}\right\Vert _{F}^{2}+\frac{\lambda}{2}\left(\left\Vert \mathbf{P}\right\Vert _{F}^{2}+\left\Vert \mathbf{Q}\right\Vert _{F}^{2}\right)+\mu\left\Vert \mathbf{B}\right\Vert _{1},\nonumber \\
\textrm{subject to}\; & \mathbf{A=B}.\label{eq:problem-formulation-admm}
\end{align}
The augmented Lagrangian of (\ref{eq:problem-formulation-admm}) is
\begin{align*}
L_{c}(\mathbf{P},\mathbf{Q},\mathbf{A},\mathbf{B},\boldsymbol{\Pi})= & \frac{1}{2}\left\Vert \mathbf{P}\mathbf{Q}+\mathbf{D}\mathbf{A}-\mathbf{Y}\right\Vert _{F}^{2}+\frac{\lambda}{2}\left(\left\Vert \mathbf{P}\right\Vert _{F}^{2}+\left\Vert \mathbf{Q}\right\Vert _{F}^{2}\right)\\
 & +\mu\left\Vert \mathbf{B}\right\Vert _{1}+\textrm{tr}(\boldsymbol{\Pi}^{T}(\mathbf{A-B}))+\frac{c}{2}\left\Vert \mathbf{A}-\mathbf{B}\right\Vert _{F}^{2},
\end{align*}
where $c$ is a positive constant. In ADMM, the variables are updated
in the $t$-th iteration as follows:
\begin{align*}
(\mathbf{Q}^{t+1},\mathbf{B}^{t+1}) & =\underset{\mathbf{Q},\mathbf{A}}{\arg\min}\; L_{c}(\mathbf{P}^{t},\mathbf{Q},\mathbf{A}^{t},\mathbf{B},\boldsymbol{\Pi}^{t}),\\
\mathbf{P}^{t+1} & =\underset{\mathbf{P}}{\arg\min}\; L_{c}(\mathbf{P},\mathbf{Q}^{t+1},\mathbf{A}^{t+1},\mathbf{B}^{t},\boldsymbol{\Pi}^{t}),\\
\mathbf{A}^{t+1} & =\underset{\mathbf{B}}{\arg\min}\; L_{c}(\mathbf{P}^{t+1},\mathbf{Q}^{t+1},\mathbf{A},\mathbf{B}^{t+1},\boldsymbol{\Pi}^{t}),\\
\boldsymbol{\Pi}^{t+1} & =\boldsymbol{\Pi}^{t}+c(\mathbf{A}^{t+1}-\mathbf{B}^{t+1}).
\end{align*}
Note that the solutions to the above optimization problems have an
analytical expression; see \cite{Mardani2013} for more details. We
set $c=100$ in the following simulations.

\begin{figure}[t]
\center \includegraphics[scale=0.515]{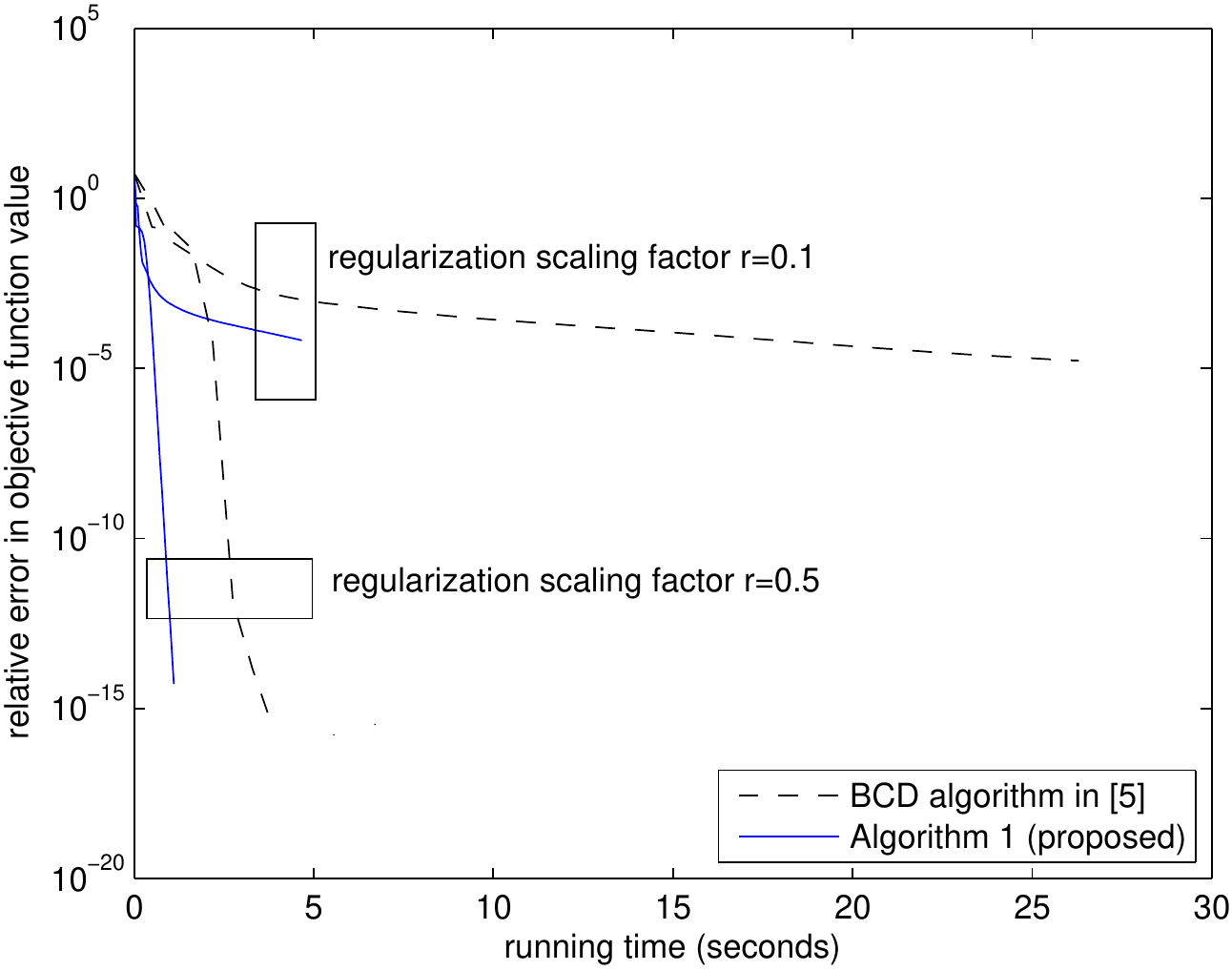}\protect\caption{\label{fig:Time}Relative error in objective function value versus
the CPU time}
\end{figure}

The simulation parameters are set as follows. $N=106$, $K=380$,
$I=380$, $\rho=3$. The elements of $\mathbf{D}$ are generated randomly
and they are either 0 or 1. The elements of $\mathbf{V}$ follow the
Gaussian distribution with mean 0 and variance $0.01$. Each element
of $\mathbf{S}$ can take three possible values, namely, -1, 0,1,
with the probability $P(S_{i,k}=-1)=P(S_{ik}=1)=0.05$ and $P(S_{ik}=0)=0.9$.
We set $\mathbf{Y}=\mathbf{PQ}+\mathbf{DS}+\mathbf{V}$, where the
elements of $\mathbf{P}$ (\textbf{$\mathbf{Q}$}) are generated randomly
following the Gaussian distribution with mean 0 and variance $100/I$
($100/K$). The sparsity regularizer $\lambda=r\left\Vert \mathbf{Y}\right\Vert $
($\left\Vert \mathbf{Y}\right\Vert $ is the spectral norm of $\mathbf{Y}$)
and $\mu=r\left\Vert \mathbf{D}^{T}\mathbf{Y}\right\Vert _{\infty}$,
where $r$ is the regularization scaling factor that is either 0.1
or 0.5.

In Figure \ref{fig:Iteration}, we show the relative error in objective
function value versus the number of iterations achieved by different
algorithms, where the optimal objective function value is computed
by running Algorithm \ref{alg:Successive-approximation-method} for
a sufficient number of iterations. As we can see from Figure \ref{fig:Iteration},
the ADMM does not always converge for both regularization parameters
$r=0.1$ and $0.5$, as the optimization problem (\ref{eq:problem-formulation-admm})
(and (\ref{eq:problem-formulation})) is nonconvex.

Note that for the BCD algorithm in Figure \ref{fig:Iteration}, all
elements of $\mathbf{S}$ are updated once, in a sequential order,
in one iteration. We can see from Figure \ref{fig:Iteration} that
the BCD algorithm converges in less number of iterations than the
proposed Algorithm \ref{alg:Successive-approximation-method}. But
the incurred delay of each iteration in the BCD algorithm is typically
very large, because all elements are updated sequentially. On the
other hand, in the proposed algorithm, all variables are updated simultaneously
and the CPU time (in seconds) needed for each iteration is relatively
small. This is illustrated numerically in Figure \ref{fig:Time},
where two regularization scaling factors are tested, namely, $r=0.1$
and $r=0.5$. We see that the improvement is notable when the regularization
parameter is small.

\section{Concluding Remarks}

In this paper, we have proposed a parallel best-response algorithm
for the nonconvex sparsity-regularized rank minimization problem.
The proposed algorithm exhibits fast convergence and low complexity,
because 1) the variables are updated simultaneously based on their
best response; 2) the stepsize is based on the exact line search and
it is performed over a differentiable function; and 3) both the best
response and the stepsize are computed by closed-form expressions.
Furthermore, the proposed algorithm has a guaranteed convergence to
the stationary point. The advantages of the proposed algorithm are
also consolidated numerically.

\end{document}